\documentstyle[11pt]{article}
\def\alt{\;{\lower 1.5pt\hbox{$<$} \above 0pt \raise 1.5pt\hbox{$\sim$}}\;}
\def\agt{\;{\lower 1.5pt\hbox{$>$} \above 0pt \raise 1.5pt\hbox{$\sim$}}\;}

\begin{document}
\vbox{
\begin{flushright}
AZPH-TH/92-29
\end{flushright}
\title{A PROPERTY OF RECOMBINATION IN POLARIZED HADRONIC TARGETS}
\author{Arjun Berera*\\
Department of Physics \\
University of Arizona \\
Tucson, Arizona 85721}
\date{}
\maketitle
\begin{abstract}
The triple gluon-ladder vertex is shown to project the outgoing gluon
in either polarization state with equal probability up to the leading
double-ln(x)ln($Q^2$) approximation.
This implies that the $Q^2$-evolution of
$\Delta G (x,  Q^2)$ is free
from recombination effects to this level of approximation.
\end{abstract}
\bigskip
\vspace{20mm}}

PACS numbers: 13.88.+e

\medskip

Phys. Rev D, in press 1994
\eject

\section {Introduction}

The A-dependence in deep inelastic scattering experiments
off nuclei has been a long studied problem. The
many levels of description of this behavior make it
a textbook example of quantum mechanical reasoning. The simple
intuitive explanation is that the nucleons at the
``front face'' of the nuclei shadow those behind them.
Quantum mechanically the
effect at very low energy is explained at a phenomenological
level by the vector-dominance-model (VDM)\cite{feynman}.  Here the
spacelike photon is supposed to have coupled to a vector meson
which then interacts with the hadrons in the nuclei.
Due to the strong interaction between the meson and nucleons
the interaction is likened once again to occur at the front
face of the nuclei.

At the microscopic level it was Zakharov
and Nikolaev\cite{zakharov} who in a classic paper
explained how the parton model accounts for the
A-dependence at large $Q^2$ through recombination of the constituents.
This description, in recent times, has been quantified by
perturbative QCD in the small-$x$ high $Q^2$ regime
\cite{mueller,qui,levin}.
The QCD picture is that the standard $Q^2$-evolved
ladders recombine via the triple gluon-ladder vertex, the basic
element of the theory.

The QCD argument has reasonable physical underpinnings
and in the large-$Q^2$ regime shows consistency
to experiment.
The question thus arises on how to put it to
further tests that reveal
qualitatively new features.  In this paper we
will show how scattering polarized protons or nuclei off
polarized nuclei could provide a simple test
of recombination effects.
It will be shown that the triple gluon-ladder vertex,
the motive of recombination, at leading-$\ln x$ order
decorrelates gluon polarization.
This means, given two gluon densities of
polarization $\alpha$, $\alpha^\prime$, if they recombine
through the triple gluon-ladder vertex, the outgoing ``recombined''
legs will be in either polarization state with
equal probability; a property which we will refer to
as disordering of polarization.

To understand the possible observable outcomes of this mechanism, let
us recall that nuclear shadowing arises from several sources, of which
recombination is only of relevance at large $Q^2$ and small-x.
At low $Q^2$, shadowing is a nonperturbative effect with recombination
playing an insignificant role.  However, the implication of
polarization disordering is that as one scales up in $Q^2$, recombination
will not modify
whatever the low energy shadowing effects are in $\Delta G$.
This statement is valid up to the leading double-ln(x)ln($Q^2$)
approximation only.
Our purpose in this paper is
only to identify this hard mechanism.  We are not claiming
that the effect is measurable.  Here, the complications arise both
from present day experimental limitations and unknown nonperturbative
inputs.  Due to the simplicity of this mechanism and its
distinct qualitative effects, we feel there may be use at some
stage for the theoretical considerations given below.

\section {Polarization Disordering}

For our calculations we will work in the light-cone frame.
For a momentum vector $p=(p_0,p_1,p_2,p_3)$ we have
$p_\pm=p_0\pm p_3$
with the
plus component $p_+$
expressed
through its momentum fraction $z_p$ as
$p_+=z_pP$.  We will use the light-cone
axial gauge so that the gluon propagator
$D_{\mu\nu}(k)$ is,
\begin{equation}
D^{ab}_{\mu\nu}(k)=\frac{\delta_{ab}}{k^2-i\epsilon}\sum^2_{\alpha=1}
e_\mu^{(\alpha)}(k)e_\nu^{(\alpha)}(k)
\end{equation}
where
\begin{equation}
e_\mu^{(\alpha)}=\left[ g_{\mu}^\nu-\frac{n_\mu k^\nu +k_\mu n^\nu}{n\cdot k}
\right]\tilde e_\nu^{(\alpha)}
\end{equation}
with
\begin{eqnarray*}
\tilde e^{(1)} & = & (0,1,0,0) \nonumber \\
\tilde e^{(2)} & = & (0,0,1,0) \nonumber \\
\end{eqnarray*}
and $n=(-1,0,0,1)$ so that $n\cdot k=k_+$.

The basis of our analysis will be
the nonlinear evolution equations involving recombination that
were obtained
by Mueller and Qiu\cite{mueller}.
The equations for the gluon density $G(x,Q^2)$
and quark density
$q(x,Q^2)$
are,
\begin{eqnarray*}
\hspace{5pc} Q^2\frac{\partial}{\partial Q^2}G(x, Q^2) & = &
	\frac{\alpha}{2\pi}\int^1_x\frac{dy}{y}P_{GG}\left(\frac{x}{y}\right)
	G(y,Q^2) \\
& - & \frac{4\pi^3}{N^2-1}\left(\frac{\alpha C_2(G)}{\pi}\right)^2
	\frac{1}{xQ^2}\int^1_x\frac{dy}{y}y^2 G^{(2)}(y,Q^2;y,Q^2)
	\hspace{3.31pc} (3a)  \\
Q^2\frac{\partial}{\partial Q^2}q(x, Q^2) & = &
	\frac{\alpha}{2\pi}\int^1_x\frac{dy}{y}2fP_{qG}\left(\frac{x}{y}\right)
	G(y,Q^2) \\
& - & \frac{2\alpha^2 f\pi}{N(N^2-1)Q^2}\left[\frac{4}{15}N^2-
	\frac{3}{5}\right]x G^{(2)}(x,Q^2;x,Q^2) \\
& + & \frac{2\alpha f}{\pi Q^2}
	\int^1_x \frac{dy}{y}\bar\gamma_{FG}\left(\frac{x}{y}\right) G_{HT}
	(y,Q^2) \hspace{10.725pc} (3b)
\end{eqnarray*}
where,
\addtocounter{equation}{1}
\begin{equation}
q(x,Q^2)=\sum^{2f}_{i=1} q_i(x,Q^2) \ ,
\end{equation}
\begin{equation}
l^2\frac{\partial}{\partial l^2}x^\prime G_{HT}=-\frac{4\pi^2}{N^2-1}
	\left(\frac{\alpha C_2(G)}{\pi}\right)^2\int^1_{x^\prime}\frac{dy}{y}
	y^2 G^{(2)}(y,\underline{l}^2;y,\underline{l}^2) \ ,
\end{equation}
\begin{equation}
\bar \gamma_{FG}(z)=- 2z+15z^2-30z^3+18z^4 \ .
\end{equation}
$G^{(2)}(x,Q^2,x^\prime,Q^{\prime 2})$
above is the two-gluon
density distribution, $C_2(G)=N$, and $f$ is the number of fermion
flavors.  Often for studies of loosely bound
systems such as a nucleus, $G^{(2)}(x,Q^2,x^\prime,Q^{\prime 2})$
is approximated as
the product of two single gluon density distributions.
Our analysis holds for the general form so we will leave
it as such.
In regards to equations (3a) and (3b),
we will only be concerned with some of their features
and will elaborate on them where needed.  For a general discussion
of these equations, the reader is referred to the original paper\cite{mueller}.

Let us now examine the triple gluon-ladder vertex.
This we define as everything except the two-gluon density
distribution in the second
term of equation (3a).
What
we will show is
that the polarization
of the gluon emerging from the bottom ladder
(labeled $m$ in figure 1) is uncorrelated to the
polarizations of the incoming gluons.
To understand this let us examine the helicity
flow at the vertices $V$ and $V^\prime$
of the triple gluon-ladder vertex in Figure 1.
The vertex contraction at $V$, for example, is by the general
formula given as,
\begin{eqnarray}
&&e^{(\alpha)\mu}(b)e^{(\alpha^\prime)\nu}(b-m)e^{(\alpha^{\prime\prime})
	\lambda}
	(m)\Gamma^{uvw}_{\mu\nu\lambda}(b,b-m,m) =  -ig c_{uvw}\nonumber\\
&& \ \ \left[e^{(\alpha)}(b) \cdot e^{(\alpha^\prime)}(b-m)[2b-m]\right.
	\cdot e^{(a^{\prime\prime})}(m) -   e^{(\alpha)}(b)\cdot
	e^{(\alpha^{\prime\prime})}(m)[b+m]\cdot e^{(\alpha^\prime)}(b-m) \\
&& \ \ + \left. e^{(\alpha^\prime)}(b-m)\cdot e^{(\alpha^{\prime\prime})}
	(m)[2m-b]\cdot e^{(\alpha)}(b) \right] \nonumber
\end{eqnarray}
The leading-$\ln x$ approximation requires retaining the
most singular term at this vertex as a function
of the longitudinal momentum.  By noting the
momentum ordering $z_b \gg z_m$,
the relevant term from (7)
is then,
\begin{equation}
-igc_{uvw}e^{(\alpha)}(b)\cdot e^{(\alpha^\prime)}(b-m)[2b-m]\cdot
e^{(\alpha^{\prime\prime})}(m)\cong -igc_{uvw}\delta_{\alpha\alpha^\prime}
\frac{2b_+}{m_+}m\cdot \tilde e^{(\alpha^{\prime\prime})}
\end{equation}
In words this says that the contraction is between the
two polarization vectors from the upper ladders
whereas the polarization vector of the $m$-line
is contracted with the momentum term from the vertex.
This means the polarization state of the $m$-gluon
is uncorrelated to the polarization states of
the two impinging gluons.
Recall now that it is the state of the m-gluon that defines the gluon
density as can be seen by inspection of the evolution equation (3a).
Thus, based on our above deduction about the vertex, we can conclude that
the polarization state in the gluon
density distribution
at the triple gluon-ladder vertex becomes disordered.

The approximation used to go from (7) to (8) is the same as what in
reference \cite{mueller} is called the leading-double-logarithmic
approximation (DLA)
in the light-cone axial gauge.  Equivalently, in the Altarelli-Parisi
formalism\cite{altarelli} the contraction given in (8) is the one
that leads to the $1/z$ term in the gluon-gluon
probability kernel.
As a check one can see from that paper that
$P_{G_+G_+}(z)=P_{G_-G_+}(z)$ at
$1/z$ order. This is indicative of the
disordering effect described above.

To make more precise our thinking, in symbolic terms
the recombination vertex is a convolution of the general
form,
\begin{equation}
G^{(2)}_{\alpha\alpha^\prime}\ast V_{\alpha\alpha^\prime\alpha^{\prime\prime}}
\end{equation}
where $G^{(2)}_{\alpha\alpha^\prime}$
is the density distribution for two gluons
in polarization states $\alpha$ and $\alpha^\prime$.
$ V_{\alpha\alpha^\prime\alpha^{\prime\prime}}$ is the triple gluon-
ladder vertex connecting the incoming polarization states
$\alpha$ and $\alpha^\prime$
to the outgoing polarization state $\alpha^{\prime\prime}$, where
$\alpha$ is defined as the outer gluon density in
Figure 1.  Our above analysis
of $V_{\alpha\alpha^\prime\alpha^{\prime\prime}}$
shows that
\begin{equation}
V_{+\alpha^\prime+}=V_{+\alpha^\prime-}=V_{-\alpha^\prime+}=V_{-\alpha^\prime
-}
\end{equation}
Since the form of the recombination term contributing
to $\Delta G$ is,
\begin{equation}
\Delta G_R=\sum_{\alpha\alpha^\prime}G_{\alpha\alpha^\prime}\ast
(V_{\alpha\alpha^\prime +}-V_{\alpha\alpha^\prime -})
\end{equation}
by using equation(10) we see that $\Delta G_R =0$.  As an aside
notice that the incoming gluons connected to the inner vertices in
Figure 1 play no role in controlling
the polarization flow across the outer vertices $V$ and $V^\prime$.

This demonstration is not complete since the diagram in
figure 1 is only one among a group of triple gluon-ladder vertices.
However Mueller and Qiu\cite{mueller} showed that by taking advantage of
the residual gauge freedom available in the axial gauge,
one can impose the prescription for
the $1/k_+$ denominator as,
\begin{equation}
\frac{1}{k_+}\equiv\frac{1}{k_+ -i\epsilon}
\end{equation}
With this choice they demonstrated
that the diagram in Figure 1 is in fact
the only one at leading-$\ln x$ order.
To simplify our discussion
we will adhere to their prescription.

At this stage we remind the reader that the polarization
vectors in equation (2) are not the same as for physical gluons
due to the longitudinal components.  However, to leading
$\ln Q^2$ order they are equivalent.  We also
clarify that we are not interpreting their relation to the physical
polarization vectors in any way differently
then as done in the Altarelli-Parisi formalism.
For this recall that the recombination calculation
is done in the probabilistic limit of the QCD equations which is
the same limit as for the Altarelli-Parisi equations.
In this
limit Gribov\cite{gribov1} has shown that the results are gauge independent
and can be obtained, in fact, by using only the two physical
transverse polarizations of the gluons.  With some
consideration one can be convinced that the transverse components
of the light-cone axial gauge polarization vectors retain
the same physical information as the physical
transverse vectors of Gribov.  The point of this discussion is
to clarify the interpretation of our findings in terms of
the physical densities that are ultimately probed.  We are simply
clarifying that the correspondence here is the obvious one
that one would naturally suspect.

The reason for using the axial gauge in recombination calculations
is due to the simplification in the number of graphs
to be evaluated, which in particular is markedly decreased
by choosing an appropriate prescription for
the residual gauge freedom.
In the original work of Altarelli-Parisi\cite{altarelli} where only
ladder diagrams were discussed, it is obvious by
inspection on a graph by graph basis that all residual
gauge choices are equivalent.  This is because no
expressions of the form $(1/ k_- )^2$
ever arise.  It is for this reason that in their derivation
the issue of residual gauge choice never came up.
To the authors knowledge, it was
Mueller and Qiu \cite{mueller} who recognized how to take advantage
of gauge prescriptions to
simplify the recombination calculation.

We now obtain from equation (3a) the evolution equation for
$\Delta G$ as,
\begin{equation}
Q^2\frac{\partial}{\partial Q^2}\Delta G(x,Q^2)=\frac{\alpha}{2\pi}
\int^1_x\frac{dy}{y}\Delta P_{GG}\left(\frac{x}{y}\right)\Delta G(y,Q^2)
\end{equation}
where
\begin{equation}
\Delta G(x,Q^2)\equiv G_+(x,Q^2)-G_-(x,Q^2)
\end{equation}
Observe that this equation has no recombination effects
in the leading ln(x)ln($Q^2$) limit.
This is a direct outcome of our basic
observation  regarding polarization flow and
is a property of perturbative QCD, free from model
dependent assumptions.

\section {Conclusion}

Recall that perturbative QCD-recombination is operative at large
$Q^2$.  To understand the implications of polarization disordering
from recombination, as given in eqs (13), let us consider
two possible low energy scenarios.  In the first case, which is in
the spirit of \cite {mueller}, suppose
shadowing effects in polarized nuclei at low-$Q^2$ are given by
some low-energy model.  Then according to eqs (13), no further
shadowing effects will be generated by QCD-evolution for
$\Delta G$.  The second example, which is in the spirit
of \cite{close}, supposes that at the low energy scale there is no
shadowing.  In their model, shadowing is soley attributed
to QCD-recombination.  Due to polarization disordering, it follows
that to leading double-ln(x)ln($Q^2$) shadowing would not be seen in
$\Delta G(x, Q^2)$ as $Q^2$ is increased.  In both examples,
nonleading gluon-gluon fusion and quark-gluon fusion would
be expected to give rise to recombination induced
shadowing.  We will not pursue that matter
at present.

\bigskip

\Large
\section {Acknowledgements}

\medskip

\normalsize

I thank Professors James D. Bjorken,
Ian Hinchliffe, Alfred H. Mueller, Jianwei Qiu, and
Mark Strikman
for helpful discussions.
Financial support was provided
by
the U. S. Department
of Energy,
Division of High Energy and Nuclear Physics.

*  Present address: Department of Physics, the Pennslyvania State University,
104 Davey Laboratory, University Park, PA 16802-6300

\end{document}